\documentclass[sigconf]{acmart}
\usepackage{placeins} 

\setcopyright{acmcopyright}
\copyrightyear{2026}
\acmYear{2026}
\acmDOI{3773078.3831785}

\acmConference[RecSys '26]{ACM Conference on Recommender Systems}{September 28--October 2, 2026}{Minneapolis, Minnesota, USA}

\usepackage{booktabs}
\usepackage{balance}
\usepackage{graphicx}
\usepackage{enumitem}

\newcount\Comments  
\newcommand{\kibitz}[2]{\ifnum\Comments=1\textcolor{#1}{#2}\fi}

\begin{document}

\title{Who Are We Recommending To? Recommender Systems in the Agentic Web}

\author{Himan Abdollahpouri}
\affiliation{%
  \institution{Spotify}
  \country{USA}
}
\email{himana@spotify.com}

\author{Kyle Kretschman}
\affiliation{%
  \institution{Spotify}
  \country{USA}
}
\email{kylek@spotify.com}

\author{Sai	Ravindranath}
\affiliation{%
  \institution{Spotify}
  \country{USA}
}
\email{saisrivatsar@spotify.com}

\author{Jackie	Doremus}
\affiliation{%
  \institution{Spotify}
  \country{USA}
}
\email{jackied@spotify.com}

\author{Mounia Lalmas}
\affiliation{%
  \institution{Spotify}
  \country{UK}
}
\email{mounia@acm.org}

\begin{abstract}

For two decades, recommender systems have been designed under the assumption that a human directly consumes each recommendation: receiving, interpreting, and acting upon it. The emergence of AI agents powered by large language models challenges this assumption. In the emerging \emph{Agentic Web} \cite{yang2025agentic}, autonomous agents increasingly act on behalf of users, e.g., browsing, comparing, negotiating, and executing transactions, raising a central question: who is the receiver of a recommendation? 
In this position paper, we argue that the recommendation paradigm is undergoing a bifurcation. In \emph{delegable} contexts, such as routine purchases, travel, and constrained transactional tasks, the primary operational consumer of recommendations is shifting from the human to the agent, requiring new optimization objectives, interaction protocols, and evaluation criteria. In \emph{experiential} contexts, such as entertainment, art, and other subjective or high-stakes choices, humans remain the final judge of relevance, though agents may assist through pre-filtering and curation. We introduce a \emph{delegation spectrum} that characterizes recommendation contexts along factors such as preference specifiability, outcome verifiability, and decision stakes, and we outline a research agenda spanning agent preference modeling, dual-audience optimization, and the emerging agent attention economy. We further discuss the implications of this shift for the design and evaluation of recommender systems.

\end{abstract}

\ccsdesc[500]{Information systems~Recommender systems}
\ccsdesc[300]{Computing methodologies~Artificial intelligence}

\keywords{Recommender systems, AI agents, agentic web, large language models, delegation spectrum, agent attention economy}

\maketitle

\section{Introduction}
\label{sec:intro}


The ACM Conference on Recommender Systems marks its twentieth year. Over this period, the field has progressed from neighborhood-based collaborative filtering~\cite{sarwar2001item} through matrix factorization~\cite{koren2009matrix} and deep neural architectures~\cite{he2017neural,cheng2016wide} to recent large language model (LLM)--augmented recommenders~\cite{wu2024survey}. Despite these advances, one assumption has remained largely unchanged: recommendations are ultimately received, interpreted, and acted upon by humans.

This assumption has shaped the entire recommender systems stack. Interfaces are optimized for human inspection through ranked lists and feeds; evaluation relies on engagement-based proxies such as clicks, dwell time, and conversions; explanations are designed for interpretability via natural language and social signals; and fairness is defined in terms of exposure, diversity, and user experience. In short, recommender systems have historically been designed around human cognition and perception.


This foundational assumption is now under pressure. The maturation of LLM-powered AI agents, i.e., autonomous software entities that perceive, reason, plan, and execute multi-step tasks on behalf of users~\cite{wang2024survey}, is contributing to what has been termed the \emph{Agentic Web}~\cite{yang2025agentic}. Unlike earlier forms of digital intermediaries, these agents do not merely surface options but can act on them. They can book flights, compare insurance policies, negotiate prices, and orchestrate complex workflows with minimal human intervention. For example, on an e-commerce platform that offers agentic AI shopping assistance, the user may ask "Buy me the best wireless earbuds under \$250 that work well with my iPhone." Or on a travel platform, the user may ask "Find the cheapest nonstop flight to Tokyo in October and book a refundable hotel near Shibuya". When an agent acts on behalf of a user, recommendations are no longer necessarily consumed by a human scanning a ranked list, but may instead be processed by a system that parses structured data, reasons over constraints, and executes transactions, potentially as part of a broader decision pipeline.


This shift does not render human-facing recommendation obsolete, but it challenges the assumption that the consumer of a recommendation is always a human and whether recommendations should always be optimized and presented based on human perception and cognition. The extent to which agents can substitute for human judgment appears \emph{context-dependent}. Tasks such as booking a flight involve well-defined, verifiable objectives that can be optimized algorithmically, whereas choices such as selecting a gift for a loved one depend on subjective, experiential factors that resist full delegation. The degree of delegation varies across decision contexts, depending on how preferences are specified, how outcomes can be evaluated, and the stakes of the decision. Between these extremes lies a broad design space in which agents and humans jointly contribute to decision-making.

\begin{figure*}[t]
  \centering
  \includegraphics[width=0.85\textwidth]{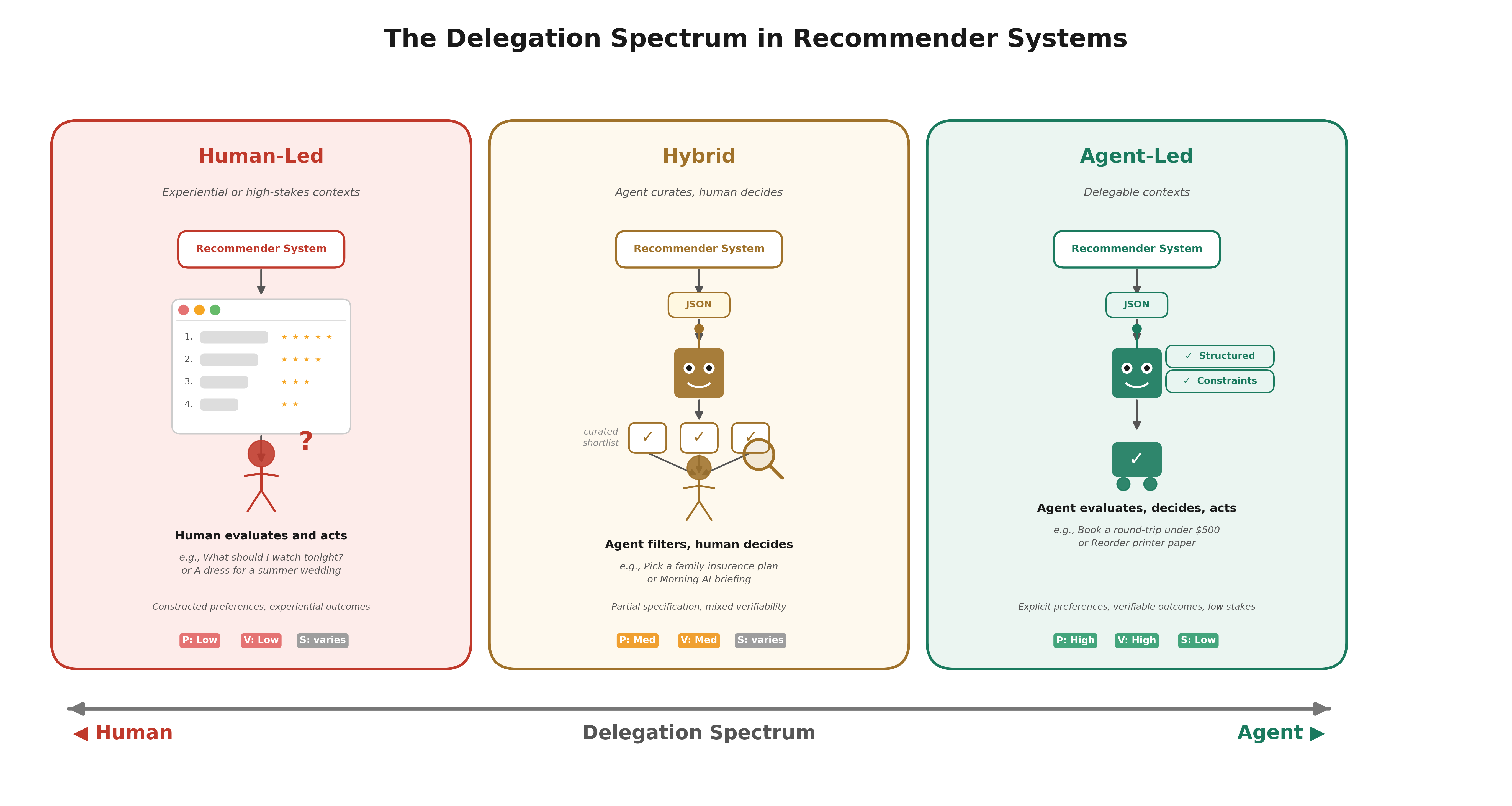}
\caption{The delegation spectrum in recommender systems, organized by decision context in the Agentic Web. From left to right: in \emph{human-led} contexts (constructed preferences, experiential outcomes), humans evaluate and act on recommendations directly; in \emph{hybrid} contexts (partial specification, mixed verifiability), agents curate a shortlist and humans make the final choice; in \emph{agent-led} contexts (explicit preferences, verifiable outcomes, low stakes), agents autonomously evaluate, decide, and transact on behalf of the user. Each panel shows the associated levels of preference specifiability (P), outcome verifiability (V), and stakes (S) that drive a context's position on the spectrum.}
  \label{fig:paradigm-shift}
\end{figure*}

In this paper, we introduce the concept of a \textbf{delegation spectrum} for recommender systems in the context of the Agentic Web (see Figure~\ref{fig:paradigm-shift}). We argue that the recommender systems community should prepare for this paradigm shift, which opens up an abundance of new research opportunities, including advances in metrics and algorithms, interface design, user experience, explainability and transparency, and multi-objective optimization. We further formalize the implications of this shift and outline a research agenda for the community.

\section{Past and Present: Two Decades of Recommending to Humans}
\label{sec:past}

Recommender systems emerged from a simple observation: as the web scaled, people could no longer reliably find relevant items through browsing alone due to what is referred to as \textit{information overload}~\cite{resnick1997recommender, aljukhadar2010information}. The field's first decade was dominated by collaborative filtering on user-based~\cite{herlocker1999algorithmic}, item-based~\cite{sarwar2001item}, and model-based variants~\cite{koren2009matrix} largely optimized for predicting human judgments and actions like ratings or clicks. The Netflix Prize exemplified this objective, focusing on minimizing root-mean-square error over a matrix of \emph{human} preferences~\cite{bennett2007netflix}.

The mobile web era brought a shift from explicit ratings to implicit signals (e.g., clicks, dwell time, scrolls, purchases) and with it, deep learning architectures capable of modeling high-dimensional behavioral patterns~\cite{covington2016deep,cheng2016wide,guo2017deepfm}. Sequential models captured temporal dynamics in user behavior~\cite{hidasi2016session,kang2018self,sun2019bert4rec}, while graph neural networks modeled social influence~\cite{wang2019neural}. At the same time, fairness, diversity, and explainability emerged as important considerations, grounded in \emph{human} perception, such as whether users can understand recommendations and whether result sets reflect meaningful diversity.

More recently, conversational and LLM-based recommenders have introduced natural-language interaction~\cite{gao2021advances,wu2024survey}, expanding how users express intent and consume recommendations. Despite these advances, the consumer of the recommendation remains a human: issuing queries, inspecting results, and making final decisions. Even the most advanced systems assume a human-in-the-loop at the point of consumption, with the recommender system responsible for ranking rather than executing decisions.


This human-centric design has been a feature rather than a limitation. However, it encodes an implicit contract: recommender systems optimize, and users evaluate. This depends on the consumer being a human. What happens when that consumer is an agent? The history of our field has been a history of the human consumer: from the user who rates items (Past), to the user who browses feeds (Present), and now to the user who may delegate tasks to agents (Future).

A parallel line of work repositions the recommender system itself as an LLM-based agent~\cite{huang2025towards}, focusing on how internal agentic capabilities such as planning, memory, and multimodal reasoning can improve recommendation quality for human consumers. Our focus is orthogonal: we ask what changes when the consumer of the recommendation is itself an agent acting on behalf of the user.

\section{The Paradigm Shift: Recommending in the Agentic Web}
\label{sec:shift}

We now turn from historical evolution to a structural shift: recommender systems are increasingly integrated into agent-driven pipelines where recommendations may be consumed and acted upon by autonomous systems.


\subsection{From Attention Economy to Agent Attention Economy}

Yang et al.~\cite{yang2025agentic} describe the evolution of the web across three eras: the PC Web (search paradigm), the Mobile Web (recommendation paradigm), and the Agentic Web (action paradigm). In the mobile era, recommender systems became a primary mechanism for matching information supply and demand, contributing to what has been termed the \emph{attention economy}~\cite{davenport2001attention}, in which platforms compete for human attention.

The Agentic Web introduces a parallel economy: the \emph{agent attention economy}, in which services, tools, and content providers compete to be selected and invoked by autonomous agents rather than clicked by humans. In this economy, recommendations serve a dual role. On one hand, they must still satisfy the latent preferences of the end user (who remains the ultimate beneficiary). On the other hand, they must be \emph{actionable} by the agent---structured, machine-readable, and responsive to constraint satisfaction and multi-step planning.

Crucially, recommendation consumption is no longer tied to a single type of consumer. In some settings, agents can fully evaluate, decide, and act on recommendations; in others, human judgment remains essential. As a result, the consumer of a recommendation is no longer fixed, but varies along a continuum between human-led and agent-led decision-making.

\subsection{The Delegation Spectrum}
\label{sec:spectrum}

We capture this variation through a \emph{delegation spectrum} that characterizes how recommendation consumption varies across decision contexts.
Not all recommendation tasks are equally suited to agent delegation. For example, booking the cheapest
nonstop flight for a recurring business trip is highly delegable, while
planning a honeymoon is not; reordering printer paper is delegable, while
choosing a gift for a close friend is not. We therefore ground delegability in three underlying factors that jointly
determine how much of a decision can be safely handed to an agent:

\begin{itemize}
  \item \textbf{Preference specifiability}: the extent to which the user's
  preferences can be articulated \emph{ex ante} as explicit constraints and
  objectives, versus being \emph{constructed} through exposure to options,
  reflection, or lived experience~\cite{bettman1998constructive}.
  Specifications like ``under \$500, nonstop, arriving before 6pm'' are
  explicit; ``a dress that feels right for my friend's wedding'' is
  constructed.

  \item \textbf{Outcome verifiability}: the extent to which the quality of
  the chosen item can be assessed against objective, machine-checkable
  criteria (price, delivery time, compatibility, specifications) rather than
  requiring subjective human experience (taste, enjoyment, emotional fit).
  An agent can verify that a battery meets a voltage spec; it cannot verify
  that a novel will move the reader.

  \item \textbf{Stakes and reversibility}: the magnitude and reversibility
  of the decision's consequences. Low-stakes or easily reversible choices
  (a \$6 pack of batteries, a streaming recommendation) tolerate agent
  errors; high-stakes or irreversible ones (medical treatment, home
  purchase, major financial commitments) demand human oversight even when
  preferences and outcomes are relatively well-specified.
\end{itemize}

\smallskip\noindent
Taken together, these yield a simple heuristic:
\emph{delegation works best when preferences are explicit, outcomes are
easy to verify, and stakes are low.} As any of the three factors weakens,
the appropriate mode shifts from \emph{autonomous execution} toward
\emph{agent-assisted curation with human final judgment}, and eventually to
\emph{human-in-the-loop} decisions where the agent's role is limited to pre-filtering. Table~\ref{tab:spectrum} illustrates the spectrum with representative decisions and their preference specifiability (P), outcome verifiability (V), stakes (S), and delegation level.

\begin{table}[t]
\caption{Representative decisions along the delegation spectrum. Position
depends on preference specifiability (P), outcome verifiability (V), and
stakes (S); domains are shown as loose clusters rather than fixed
categories.}
\label{tab:spectrum}
\footnotesize
\begin{tabular}{@{}p{2.4cm}p{1.4cm}p{0.9cm}p{2.4cm}@{}}
\toprule
\textbf{Example decision} & \textbf{Delegation} & \textbf{P/V/S} & \textbf{Primary consumer} \\
\midrule
``Reorder my usual printer paper''            & Full         & H/H/L & Agent \\
``Book a nonstop round-trip under \$500''     & High         & H/H/L & Agent \\
``Find a gift for my sister's birthday''      & Low          & L/L/M & Human \\
``Plan our honeymoon in Japan''               & Minimal      & L/L/H & Human (agent assists) \\
``Pick a family health-insurance plan''       & Medium--high & M/M/H & Agent + human \\
``Morning briefing on AI policy''             & Medium       & M/M/L & Human (agent curates) \\
``What should I watch tonight?''              & Low          & L/L/L & Human \\
``A dress for a summer wedding''              & Minimal      & L/L/L & Human \\
\bottomrule
\end{tabular}
\end{table}

At one extreme, \textbf{fully delegable} decisions (explicit preferences,
verifiable outcomes, low stakes) let an agent evaluate, compare, and transact autonomously; the recommender's role shifts from influencing human choice to supporting machine-driven decision-making. At the other
extreme, \textbf{experiential or high-stakes} decisions keep the human as the final arbiter, with the agent limited to pre-filtering or contextualization. Between these poles lies a large \textbf{hybrid} region
where agents narrow the candidate set and may execute routine sub-actions, but human judgment remains necessary for choices involving subjectivity, risk, or personal values. This intermediate region is of particular
importance for system design, as it requires explicit coordination between
agent-driven filtering and human evaluation.

This variation along the delegation spectrum has direct implications for how recommender systems should be designed and evaluated.
\subsection{Implications of Agent-Consumed Recommendations}
\label{sec:agent_consumer}

As recommendation consumption ranges from human-led to hybrid to agent-led contexts, core design dimensions must adapt accordingly:
(i)~\emph{interaction format}, since ranked lists and visual cards must be complemented by structured, machine-readable outputs with attribute-level scores, constraint annotations, and provenance metadata~\cite{yang2025agentic};
(ii)~\emph{preference representation}, since implicit behavioral signals must be complemented by explicit, compositional specifications of goals, constraints, and policies;
(iii)~\emph{optimization objectives}, which shift from engagement proxies toward task completion under user-specified constraints;
(iv)~\emph{explanation and transparency}, which must evolve from natural-language rationales designed for human users~\cite{tintarev2010designing} to machine-interpretable justifications that downstream reasoning and constraint-checking can consume;
(v)~\emph{temporal dynamics}, as agents operate in tight retrieve-evaluate-execute loops that demand low-latency, high-throughput pipelines; and
(vi)~\emph{monetization}, since sponsored placements that rely on human attention lose effectiveness when a human never sees the recommendation slate.
We develop each of these as a research direction in Section~\ref{sec:agenda}.

\section{Research Directions}
\label{sec:agenda}

The shift from human to agent consumers opens a set of research
directions for recommender systems. We group them into two themes. The first four extend established recommender systems questions to a setting in which the consumer of the recommendations might be an agent either in full autonomy or in some capacity conducting pre-filtering and contextualization of the recommendations for the end user. The last two identify
new risks that arise specifically because an agent now mediates the transaction.

\paragraph{Extensions to existing recommender systems.}

\begin{enumerate}[leftmargin=*,itemsep=2pt]
\item \textbf{Dual-Audience Recommendation.} Recommender systems are unlikely to transition abruptly from human-facing to agent-facing use. A central near-term challenge is the design of systems that support both audiences: producing human-interpretable outputs for users who wish to browse, and machine-actionable representations for agents executing delegated tasks, as discussed in Section~\ref{sec:agent_consumer}. This requires formalizing the notion of actionability for agent consumers, including the representation of structured attributes, constraint satisfaction, and information about data origin and lineage to support downstream reasoning.

\item \textbf{Agent Preference Modeling. } Agents can also operate under fundamentally different cognitive constraints than humans: they have persistent memory, fast compute cycles, and the ability to maintain and update high-dimensional representations over time. While humans cannot reliably specify or manage many constraints, agents can aggregate past interactions, contextual signals, and inferred preferences into rich, structured representations. This raises the challenge of how recommender systems can ingest and effectively leverage such representations. Existing systems are typically designed to operate on implicit behavioral signals, whereas agent-mediated settings require handling explicit, compositional preference specifications, potentially transmitted through protocols such as MCP (Model Context protocol)~\cite{hou2025model} or A2A (Agent2Agent) ~\cite{ray2025review}. A key research challenge is the translation between preference representations maintained within the agent and the feature spaces used by recommender systems.

\item \textbf{Evaluation Beyond Engagement. } The dominant evaluation metrics in recommender systems (i.e., precision, recall, NDCG, click-through rate) primarily measure human engagement. In agent-mediated settings, evaluation must instead focus on task outcomes, i.e., whether recommendations enable the agent to satisfy user-specified goals and constraints. This includes assessing whether selected items meet objective criteria (e.g., budget, timing, or quality constraints), measuring regret, or how they compare to available alternatives under those criteria. Such a perspective connects to emerging work on outcome-oriented evaluation~\cite{askarbekuly2025outcome} and goal-conditioned recommendation, but also requires new benchmark designs in which the ``user'' is represented as an agent with explicit objectives and constraints. In hybrid settings, evaluation must further capture interaction dynamics, including override rates, intervention frequency, and time-to-decision, reflecting how effectively systems balance automation with human control.

\item \textbf{Monetization.} Beyond adversarial concerns, the shift to agent-mediated consumption raises new questions for monetization. Recommendation-driven advertising has historically relied on human attention, for example through sponsored placements and visually integrated content. In agent-facing settings, where recommendations may not be directly observed by users, monetization mechanisms must instead rely on signals that are interpretable by agents, such as structured attributes or service guarantees. In hybrid settings, where agents filter candidates and humans make final decisions, monetization must account for both audiences. This introduces challenges in distinguishing intrinsic quality signals from paid influence embedded in machine-readable representations, as well as in designing mechanisms for disclosure and auditing in agent-mediated interactions.
\end{enumerate}

\paragraph{New risks from agent mediation.}
\begin{enumerate}[resume,leftmargin=*,itemsep=2pt]

\item \textbf{Trust \& Accountability.} Delegating decisions to agents introduces new challenges for trust and accountability in recommender systems. When agent actions are based on recommendations, errors may arise from multiple sources, including biased data, model limitations, or misalignment between user intent and agent interpretation. This raises the question of how responsibility and accountability can be characterized across system components. From a system design perspective, this setting requires mechanisms for \emph{agent-mediated trust}. This includes modeling and estimating delegation confidence, i.e., the likelihood that an agent's action satisfies user preferences and constraints; providing auditable decision traces that capture the sequence of recommendations, intermediate reasoning steps, and executed actions; and designing fallback strategies that reintroduce human oversight when confidence is low or uncertainty is high. These challenges connect to broader questions of interpretability, reliability, and alignment in sequential decision-making systems~\cite{amodei2016concrete}, and require new methods for representing uncertainty, attributing outcomes to system components, and supporting user control in agent-mediated interactions~\cite{chan2024visibility}.

\item \textbf{Manipulation.} As services compete for agent invocation rather than human interaction, new forms of manipulation may emerge. Analogous to search engine optimization, actors may attempt to influence agent decisions by optimizing for machine-consumable signals, for example through inflated or strategically structured metadata, adversarial prompt injection, or synthetic content designed to affect LLM-based reasoning. These behaviors extend established challenges in recommender systems, including shilling attacks~\cite{lam2004shilling}, into the agent-mediated setting.

\end{enumerate}


\section{Conclusion}
\label{sec:conclusion}

For two decades, recommender systems have been built on the premise that recommendations are received, assessedm and consumed by humans. The emergence of the Agentic Web challenges this premise, not uniformly, but along a context-dependent delegation spectrum that ranges from agent autonomy in transactional contexts to essential human involvement in experiential ones. This shift does not render existing approaches obsolete; rather, it introduces a complementary design space in which the consumer of a recommendation may be an autonomous agent with distinct interaction patterns, preference representations, and evaluation criteria.


Evidence of this shift is already emerging, as platforms expose agent-facing interfaces and users increasingly delegate routine decisions to AI systems. In this context, recommender systems must support a dual-audience setting, serving both the humans who ultimately benefit from recommendations and the agents that act on their behalf. Addressing this setting raises new challenges in representation, optimization, evaluation, and system design, and defines a broad research agenda for the field.

\balance


\bibliographystyle{ACM-Reference-Format}
\balance
\bibliography{references}

\end{document}